# Mass Mixing, the Fourth Generation, and the Kinematic Higgs Mechanism

Alexander N. Jourjine[1]

Max-Planck Institute for Physics of Complex Systems
Nöthnitzer Strasse 38, 01187 Dresden, Germany

**Abstract**

We describe how to construct explicit chiral fermion mass terms using Dirac-Kähler (DK) spinors. Classical massive DK spinors are shown to be equivalent to four generations of Dirac spinors with equal mass coupled to a background $U(2,2)$ gauge field. Quantization breaks $U(2,2)$ to $U(2) \times U(2)$, lifts mass spectrum degeneracy, and generates a non-trivial CKM mixing.

## 1. Introduction

Using DK spinors to describe generations of lepto-quarks is an old idea. It was actively discussed about thirty years ago because a single DK spinor can describe a multiplet of four Dirac spinors. After the connection between spinors and differential forms was discovered in [1] and DK spinors came to physicists attention within the context of lattice fermion doubling problem [2, 3] it has been proposed in [4, 5]. These attempts to solve the generation puzzle were not successful. For once it was not clear how to extract Dirac spinors from DK spinors in a covariant way in the presence of gravity [6, 7]. Further, the standard quantization of DK spinors leads to negative norm states [8]. Lastly, at the time it was generally believed that there exist only three generations. Since then evidence has accumulated that the electroweak precision data does not exclude the fourth generation and that its existence can be beneficial, for example, for addressing such problems as baryogenesis, the coupling constant unification, and Higgs via quark condensate [9, 10]. Yet the generation structure of the SM still remains a mystery. Also the fermion mass generation via Higgs mechanism remains puzzling [11]. In this Letter we re-examine the use of DK spinors for building of four-family models and show that they can be useful for explaining the fermion and electroweak gauge field mass generation. We also show that in the DK gauge theory fermion and gauge field mass generation can be treated as two independent issues. Thus the DK theory provides at least a partial answer to the puzzle that was emphasized in [11]. Unexpectedly, a CKM-like mixing appears as a result of quantization. Here we explore only the general theoretical issues, leaving detailed comparison with the experimental data for another publication.

The Letter is organized as follows. In Section 2 we describe the covariant extraction of the Dirac degrees of freedom from DK spinors. In Section 3 we describe the general differences between the Dirac gauge theory and the DK gauge theory. In Section 4 we apply the general results to the quark sector of 4-generation DK extension of the SM. There we write down explicit dimension three mass terms that do not involve the use of a physical Higgs field. A surrogate Higgs field appears, however, and mimics the standard Higgs mechanism. Quantization and its effects on the mass spectrum and mass mixing are described in Section 5. In Section 6 we apply the results of Sections 2, 3 to the problem of electroweak gauge field mass. Section 7 is a summary.

## 2. DK Spinors in Spinbein Representation

Given a pseudo-Riemannian manifold $M$, a DK spinor is defined as an inhomogeneous differential form $f$ with values in the Lie algebra of the internal symmetry group. Let $*$ be the Hodge star map in the space of differential forms on $M$, $d$ the exterior derivative, and $\delta$ its adjoint, $\delta = -*d*$. One defines massless on-shell DK spinors as solutions of $(d-\delta)f = 0$. The relation between differential forms and spinors on $M$ is transparent in the special basis in the space of differential forms [3]. Given an orthonormal frame one-forms $e^a$ and tangent space $\gamma$-matrices $\gamma^a$, it is defined as $Z = \sum (1/p!) \gamma_{a_p} \cdots \gamma_{a_1} e^{a_1} \cdots e^{a_p}$. In this basis $f$ can be written as $f = Tr(\Psi Z)$. Under the general coordinate transformations $\Psi$ transforms as a scalar, while under local Lorentz frame rotations $e^a \to \Lambda^a_b e^b$ it transforms as a multiplet of integer-valued spin fields according to

$$\Psi(x) \to S(\Lambda) \Psi(x) S(\Lambda)^{-1}, \tag{1}$$

where $S$ is a spinor representation of $SO(1,3)$. Equation $(d-\delta)f = 0$ can be written as

$$\gamma^a e^\mu_a (i\partial_\mu \Psi + \omega_{\mu bc}[\sigma^{bc}, \Psi]) = 0, \quad \sigma^{bc} = (i/4)[\gamma^b, \gamma^c], \tag{2}$$

where $\omega_{ab} = \omega_{\mu ab} dx^\mu$, are the spin connection one-forms [4, 12]. When $M$ is flat (2) reduces to four Dirac equations. Therefore, one can identify the four columns of $\Psi$, the left invariant ideals of Clifford algebra [6], with a generation multiplet of four Dirac spinors. However, in the presence of gravity (2) differs from the equation satisfied by the Dirac spinors and such identification is possible only for flat $M$.

We now describe a generally covariant way of identifying the Dirac degrees of freedom in DK spinors. Assuming that $M$ admits a spin structure, we take two sets of Dirac spinors $\xi^A$ and $\eta^A$, $A = 1,\ldots,4$, and introduce a decomposition

$$\Psi_{\alpha\beta} = \xi^A_\alpha \overline{\overline{\eta}}^A_\beta, \tag{3}$$

where we defined $\overline{\overline{\eta}}^A$, the DK conjugate of $\eta^A$, by $\overline{\overline{\eta}}^A = \Gamma^{AB}\overline{\eta}^B$, $\overline{\eta}^B = \eta^{B*}\gamma^0$, and denoted $\Gamma \equiv diag(1, 1, -1, -1)$. Spinors $\eta^A$ are required to be orthonormal in the sense that

$$\overline{\overline{\eta}}^A_\alpha \eta^B_\alpha = \delta^{AB}, \quad \eta^A_\alpha \overline{\overline{\eta}}^A_\beta = \delta_{\alpha\beta}, \tag{4}$$

where the second equation follows from the first. The set of $\xi^A$ contains the same number of degrees of freedom as $\Psi$ and thus can represent all of its physical degrees of freedom. We take $\xi^A$ to be anti-commuting. Following [7] we assume that $\Psi$ are also anti-commuting. Therefore, $\eta^A$ have to be commuting and dimensionless. They shall represent non-physical degrees of freedom of $\Psi$. Note that, since (4) contains $\gamma^0$ and $Tr\gamma^0 = 0$, for space-times with Minkowski signature we cannot define orthonormality as $\overline{\eta}^A_\alpha \eta^B_\alpha = \delta^{AB}$.

Decomposition (3) is similar to that of metric via a vierbein. Hence, we shall call it the spinbein decomposition. A set of $\eta^A$ satisfying (4) shall be called a spinbein. It is clear that the decomposition (3) is generally covariant. The notion of spinbein is not entirely new.

Spinbein decomposition with unnormalized constant spinbeins has been used in [13]. In the phase space spinbeins appear as the normalized solutions of the Dirac equation.

Representation (3, 4) is trivially invariant with respect to simultaneous local $U(2,2)$ transformations of $\xi^A$ and $\eta^A$

$$\xi^A \to V^{AB}\xi^B, \quad \eta^A \to V^{AB}\eta^B, \quad \Gamma V^+ \Gamma V = 1. \tag{5}$$

Therefore, the coordinate dependence of $\eta^A$ can be eliminated and we can choose a spinbein with $\partial_\mu \eta_0^A = 0$. This condition also fixes the $U(2,2)$ gauge. Borrowing terminology from the Higgs mechanism, we shall call such a spinbein (gauge) the unitary spinbein (gauge). Choosing a unitary gauge breaks local $U(2,2)$ to global $U(2,2)$ and transfers all physical degrees of freedom of $\Psi$ to the multiplet of four Dirac spinors $\xi^A$. It should be noted that the related group $SU(2,2)$ appears in many unrelated subjects, such as supersymmetry, twistor theory, and ADS-CFT [14].

We now restrict ourselves to Minkowski space-time. Action for free DK spinors is

$$S = \int d^4x \, Tr\, \overline{\overline{\Psi}}(i\partial)\Psi, \tag{6}$$

where trace is over the Dirac indices of $\Psi$ and we defined the DK conjugate of DK spinor $\Psi$ as $\overline{\overline{\Psi}} \equiv \gamma^0 \Psi^+ \gamma^0$. Notation $\overline{\overline{\Psi}}$ should not be confused with the notation for the DK conjugate of the multiplet of Dirac spinors $\xi^A$: $\overline{\overline{\xi}}^A \equiv \Gamma^{AB}\overline{\xi}^B$. From (3, 4) we obtain

$$S = \int d^4x \, \overline{\overline{\xi}}^A(i\partial\,\delta^{AB} + Y^{AB})\xi^B, \qquad Y_\mu^{AB} \equiv i(\partial_\mu \overline{\overline{\eta}}^B \eta^A). \tag{7}$$

Spinbein orthonormality implies that $Y$-field is a $U(2,2)$ background gauge field coupled to $\xi^A$ with the unit coupling constant, for under (5) $Y_\mu \to VY_\mu V^{-1} + iV\partial_\mu V^{-1}$. The curvature of $Y_\mu$ vanishes identically, because $F_{\mu\nu} = \partial_\mu Y_\nu - \partial_\nu Y_\mu - i[Y_\mu, Y_\nu] \equiv 0$. Thus, action (7) describes a multiplet of four Dirac spinors coupled to a $U(2,2)$ gauge field with zero curvature. Equations of motion that follow from (7) show that spinbein is not a dynamical quantity. It is easy to see that the spinbein equations of motion are simply the covariant current conservation

$$\partial_\mu J^\mu + i[J^\mu, Y_\mu] = 0, \, J_\mu^{AB} \equiv \left(\overline{\overline{\xi}}^B \gamma_\mu \xi^A\right).$$

Chiral or Majorana DK spinors are defined by imposing the usual constraints on $\xi^A$. Neither the chirality nor Majorana condition can be meaningfully imposed on the spinbeins. The former leads to identically vanishing action, while for the latter (4) has no solutions for commuting $\eta^A$, because the charge conjugation matrix is anti-symmetric.

When $\xi^A$ and $\eta^A$ transform as Dirac spinors, $\Psi$ in (3) transforms as a DK spinor in (1). This clarifies the spin-statistics puzzle about DK spinors [3, 4, 5]. While $\Psi$ transforms as a multiplet of spin zero and one fields, its physical degrees of freedom transform as spin 1/2 fields. Although (3) is generally covariant, the spinbein fixing condition $\partial_\mu \eta_0^A = 0$ is not. Therefore, when gravity is present the definition of Dirac degrees of freedom in a DK spinor is observer dependent. This is not so surprising, since the Unruh and Hawking effects show that in the presence of gravity or in accelerated reference frames the Fock space is observer dependent.

### 3. DK Gauge Theory

Gauging action (6) brings novel features, because in addition to gauging the derivatives acting on Dirac spinors we can gauge the derivatives acting on the spinbein. As an example of this double gauging consider DK fields $\Psi_i, \Phi_p$ and their spinbein decompositions

$$\Psi_i = \xi_i \bar{\bar{\eta}}, \qquad \bar{\bar{\eta}} \eta = 1, \qquad \Phi_p = \varsigma \bar{\bar{\theta}}_p, \qquad \bar{\bar{\theta}}_p \theta_p = 1. \tag{8}$$

We now assume that Dirac spinors $\xi_i$ and spinbein $\theta_p$ transform in representations of internal symmetry groups $G$ and $G'$, respectively, while Dirac spinor $\varsigma$ and spinbein $\eta$ are singlets

$$\xi \to T(G)\xi, \quad \Psi \to T\Psi, \quad \theta \to S(G')\theta, \quad \Phi \to \Phi S^+. \tag{9}$$

The Lagrangian for two free massless DK fields (8) may be taken as

$$\mathcal{L} = Tr\, \bar{\bar{\Psi}}(i\partial)\Psi + Tr\, \bar{\bar{\Phi}}(i\partial)\Phi. \tag{10}$$

$\mathcal{L}$ is invariant with respect to global $G \times G'$ and with respect to local $U(2,2) \times U(2,2)$. Using spinbein decompositions and requiring invariance with respect to local $G \times G'$ transformations results in the minimally gauged Lagrangian that contains gauge fields $A_\mu$ and $B_\mu$

$$\mathcal{L} = \bar{\bar{\xi}}(i\mathbb{D})\xi + \bar{\bar{\varsigma}}(i\mathbb{D})\varsigma, \tag{11}$$

$$(i\mathbb{D}\xi)_i^A \equiv \{i\partial \delta_{ij} \delta^{AB} + \delta_{ij} Y^{AB} + gA_{ij}\delta^{AB}\}\xi_j^B, \quad Y_\mu^{AB} = i(\partial_\mu \bar{\bar{\eta}}^B \eta^A), \tag{12}$$

$$(i\mathbb{D}\varsigma)^A \equiv \{i\partial \delta^{AB} + \tilde{Y}^{AB} - g'\bar{\bar{\theta}}_p^B B_{pq} \theta_q^A\}\varsigma^B, \quad \tilde{Y}_\mu^{AB} = i(\partial_\mu \bar{\bar{\theta}}_p^B \theta_p^A). \tag{13}$$

In the unitary gauge (11) reduces to

$$\mathcal{L} = \bar{\bar{\xi}}_i^A i(\partial \delta_{ij} - igA_{ij})\xi_j^A + \bar{\bar{\varsigma}}^A i(\partial \delta^{AB} - ig'X^{AB})\varsigma^B. \tag{14}$$

The first term of (14) describes generation-diagonal interaction of four generations of the gauge group multiplet $\xi_i^A$ with gauge field $A_\mu$. The second term describes four generations of gauge singlets $\varsigma^A$ interacting with the gauge field $B_\mu$. The interaction is mediated via gauge field $X_\mu$ in the generation space. In the unitary gauge it reduces to $X_\mu^{AB} = -\bar{\bar{\eta}}_p^B B_{\mu pq} \eta_q^A$. For Abelian $G'$ gauge field $X_\mu$ is always generation-diagonal, since then $X_\mu^{AB} = -B_\mu \delta^{AB}$.

It should be noted that the equations of motion (2) for DK spinor field in the presence of gravity can be obtained by using the double gauging of the spinbein decomposition (3) with respect to local frame rotations in (1) by gauging the diagonal subgroup of $G \times G'$ with $G = G' = SO(1,3)$. Another important example of the double gauging procedure can be found in [12]. There, for arbitrary space-time dimension, the factors $G = G' = SO(1,n)$ are gauged independently with the second factor promoted into a $G' = SO(n+1)$ gauge group using analytical continuation in the group parameter space.

The gauged spinbein $\theta_p$ has somewhat different properties then the ungauged spinbein $\eta$. Although $\widetilde{Y}_\mu^{AB}$ still transforms as a $U(2,2)$ gauge field, $\delta_{\alpha\beta}$ in the second equation in (4) is replaced by $P_{pq}$ defined by $P_{pq} \equiv \theta_p^A \overline{\overline{\theta}}_q^A$, with $P_{pq} P_{qr} = P_{pr}$, $\overline{\overline{\theta}}_p P_{pq} = \overline{\overline{\theta}}_q$, $P_{pq} \theta_q = \theta_p$.

Below we shall be particularly interested in the case of factorizable spinbein, defined by

$$\theta_p = \varphi_p \hat{\theta}, \qquad \varphi_p^* \varphi_p = 1, \qquad \overline{\overline{\hat{\theta}}} \hat{\theta} = 1 \tag{15}$$

where $\varphi$ is a gauge multiplet of scalars and $\hat{\theta}$ is an ungauged orthonormal spinbein. For factorizable spinbeins $\widetilde{Y}$ decomposes into generation-diagonal component with non-zero curvature of an Abelian gague field and a generation-mixing component with zero curvature, while the equations of motion reduce to a non-dynamical constraint

$$J^{\mu AA}(\delta_{pq} - \varphi_p \varphi_q^*) \partial_\mu \varphi_q = 0. \tag{16}$$

Therefore, also a factorizable gauged spinbein is not a dynamical quantity and all physical degrees of freedom of a gauged DK field can still be moved to the Dirac spinor multiplet of its spinbein decomposition. Note that (16) is trivially satisfied for $\varphi = const$.

### 4. Chiral Fermion Mass Terms

In the SM the left- and the right-handed lepto-quarks transform in different representations of $SU_L(2)$. As a result, dimension three gauge invariant mass terms for chiral fermions are not allowed. In the DK gauge theory they are allowed. For example, consider local $G = SU_C(3) \times SU_L(2) \times U_Y(1)$, global $G' = SU_L(2) \times U_Y(1)$, and three DK fields $\Psi$, $\Phi_u$, $\Phi_d$ with decompositions

$$\Psi = Q \overline{\overline{\eta}}, \qquad \Phi_u = u \overline{\overline{\chi}}, \qquad \Phi_d = d \overline{\overline{\theta}}. \tag{17}$$

Here the left-handed $Q$ and the right-handed $u, d$ quark fields are four generations of Dirac spinors transforming in the fundamental representation of $SU_C(3)$, while $\eta, \chi$, and $\theta$ are spinbeins[2]. Additionally, we shall take $Q, \chi, \theta$ as $SU_L(2)$ doublets with gauge transforms

$$Q \to TQ, \; \chi \to T\chi, \; \theta \to T^*\theta, \qquad \Psi \to T\Psi, \Phi_u \to T^*\Phi_u, \Phi_d \to T\Phi_d \tag{18}$$

and $u, d, \eta$ as $SU_L(2)$ singlets. The following Lagrangian with chiral DK fields is globally $SU_L(2)$ invariant

$$\begin{aligned}\mathcal{L} = Tr\big[\overline{\overline{\Psi}}(i\partial)\Psi + \overline{\overline{\Phi}}_u(i\partial)\Phi_u + \overline{\overline{\Phi}}_d(i\partial)\Phi_d\big] - \\ - m_u Tr\big[\overline{\overline{\Psi}}(i\sigma^2)\Phi_u - \overline{\overline{\Phi}}_u(i\sigma^2)\Psi\big] - m_d Tr\big[\overline{\overline{\Psi}}\Phi_d + \overline{\overline{\Phi}}_d\Psi\big],\end{aligned} \tag{19}$$

where $m_u, m_d$ are bare masses for up and down quarks. For factorizable spinbeins (18) we can take

---
[2] If we assume that $Q, u, d$ are $SU(3)$ singlets the following discussion describes lepton multiplets with Dirac mass term for neutrinos.

$$\chi_i = \varphi_i \hat{\chi}, \qquad \theta_i = \varphi_i^* \hat{\theta}, \qquad \varphi_i^* \varphi_i = 1, \tag{20}$$

where $\varphi_i$ is a $SU_L(2)$ doublet of scalars.

The field $\varphi_i$ in (20) is non-dynamical. Hence, there are no radiative corrections. However, this does not mean that it is not determined in any way. To define the physical Dirac degrees of freedom, that is to compute S-matrix scattering amplitudes, it has to be constant and constrained to lie on a 4D sphere as (20) shows. However, as in the conventional Higgs mechanism, its exact direction is not important, since in the unitary gauge the theory has a residual global $SU(2)$ invariance. Thus in the end the field is fixed and there is no arbitrariness.

In the unitary gauge for spinbeins (20) we obtain

$$\begin{aligned}
\mathcal{L} &= \mathcal{L}_\mathcal{K} + \mathcal{L}_\mathcal{M}, \\
\mathcal{L}_\mathcal{K} &= \overline{\overline{Q}}_i^A (i\partial) Q_i^A + \overline{\overline{u}}^A (i\partial) u^A + \overline{\overline{d}}^A (i\partial) d^A, \\
\mathcal{L}_\mathcal{M} &= -m_u \left( \mathcal{M}_u^{AB} \overline{\overline{Q}}_i^A (i\sigma^2) \varphi_i^* u^B - \tilde{\mathcal{M}}_u^{AB} \overline{\overline{u}}^B (i\sigma^2) \varphi_i Q_i^A \right) - m_d \left( \mathcal{M}_d^{AB} \overline{\overline{Q}}_i^A \varphi_i d^B + \tilde{\mathcal{M}}_d^{AB} \overline{\overline{d}}^A \varphi_i^* Q_i^B \right), \\
\mathcal{M}_u^{AB} &= \overline{\overline{\hat{\chi}}}^B \eta^A, \ \tilde{\mathcal{M}}_u^{AB} = \overline{\overline{\eta}}^B \hat{\chi}^A, \qquad \mathcal{M}_d^{AB} = \overline{\overline{\hat{\theta}}}^B \eta^A, \qquad \tilde{\mathcal{M}}_d^{AB} = \overline{\overline{\eta}}^B \hat{\theta}^A.
\end{aligned} \tag{22}$$

When $\hat{\chi} = \hat{\theta}$ the up and down quark mass matrices are equal. In general $\hat{\chi} \neq \hat{\theta}$ and $\mathcal{M}_u^{AB}$, $\mathcal{M}_d^{AB}$ are independent from each other. We recognize that in (21) the doublet $\varphi$ plays the role of the Higgs field of the SM, while the eigenvalues of $m_u \mathcal{M}_u^{AB}$, $m_d \mathcal{M}_d^{AB}$ play the role of the Yukawa couplings for massless up and down quarks.

If we choose $\varphi_1 = 0$, $\varphi_2 = 1$ then the gauged version of (21) reduces to action for four generations of two multiplets of massive Dirac spinors with masses $m_u^A$, $m_d^A$, $A = 1,\ldots,4$ that have to be determined by diagonalizing the mass terms

$$\mathcal{L} = \overline{\overline{Q}}_i^A (i\mathbb{D}) Q_i^A + \overline{\overline{u}}^A (i\mathbb{D}) u^A + \overline{\overline{d}}^A (i\mathbb{D}) d^A - \left( m_u \overline{\overline{Q}}_1^A \mathcal{M}_u^{AB} u^B + m_d \overline{\overline{Q}}_2^A \mathcal{M}_d^{AB} d^B + c.c. \right). \tag{23}$$

Spinbein orthonormality implies $\mathcal{M}_{u,d} \in U(2,2)$: $\mathcal{M}_{u,d} \tilde{\mathcal{M}}_{u,d} = 1$. On classical level we can use invariance of $\mathcal{L}_\mathcal{K}$ in (21), with respect to two global $U(2,2)$ factors to absorb $\mathcal{M}_{u,d}$ in the definitions of $u$ and $d$ and fermion masses are generation independent $m_u^A = m_u$, $m_d^A = m_d$.

Note that the background $U(2,2)$ gauge fields that appears after spinbein decomposition in (17) are also non-dynamical. However, it does not mean they have no physical consequence. Their physical consequence is that they introduce the spinbein bases, which in the unitary gauge define the content of the physical Dirac spinor degrees of freedom within the DK fields in (17, 20). It is exactly the mismatch between the three spinbein bases of these DK fields that defines the physical mass spectrum of the classical or quantum DK theory.

## 5. DK Spinor Quantization and Mass Mixing

Let us now consider quantization of a single massive DK field $\Phi$ with action

$$S = \int d^4x \ \Gamma^{AB} \overline{\xi}^A (i\partial - m) \xi^B. \tag{24}$$

The standard Dirac spinor quantization procedure [15] for (24) results in an unbounded Hamiltonian and thus no vacuum state can be defined [8]. The solution to this problem is to assign the creation and annihilation operators for $A = 3, 4$ opposite to the assignment for $A = 1, 2$. The expansion of massive $\Phi$ in terms of creation and annihilation operators then is

$$\xi^A(x) = \int \frac{d^3k}{(2\pi)^3} \frac{m}{k_0} \sum_{\alpha=1,2} \left[ b_\alpha^A(\vec{k}) u^{(\alpha)}(\vec{k}) e^{-ikx} + d_\alpha^{A+}(\vec{k}) v^{(\alpha)}(\vec{k}) e^{ikx} \right], \quad A = 1, 2, \quad (25)$$

$$\xi^A(x) = \int \frac{d^3k}{(2\pi)^3} \frac{m}{k_0} \sum_{\alpha=1,2} \left[ b_\alpha^{A+}(\vec{k}) u^{(\alpha)}(\vec{k}) e^{-ikx} + d_\alpha^A(\vec{k}) v^{(\alpha)}(\vec{k}) e^{ikx} \right], \quad A = 3, 4, \quad (26)$$

$$\bar{u}^{(\alpha)}(\vec{k}) u^{(\beta)}(\vec{k}) = \delta^{\alpha\beta}, \qquad \bar{v}^{(\alpha)}(\vec{k}) v^{(\beta)}(\vec{k}) = -\delta^{\alpha\beta}, \qquad \bar{u}^{(\alpha)}(\vec{k}) v^{(\beta)}(\vec{k}) = 0. \quad (27)$$

where $k_0 = \sqrt{m^2 + \vec{k}^2}$, and $u^{(\alpha)}(\vec{k})$, $v^{(\alpha)}(\vec{k})$, $\alpha, \beta = 1, 2$, form the standard spinor basis for the normalized positive and negative energy solutions of the Dirac equation. Using (25-27) we obtain a Hamiltonian that is bounded from below and the vacuum state is well-defined. Note that $u^{(\alpha)}$, $v^{(\alpha)}$ in fact form a spinbein: $u^A = (u^{(\alpha)}, v^{(\beta)})$ satisfies (4).

At this point it is useful to review the final form of the theory as applied to the quark sector of 4 generation extension of the SM. In Section 3 we derived in the general form a representation of the field content of the gauged DK spinors in terms of gauged Dirac spinors. The representation shows exactly how DK field spinbein decomposition generates the conventional gauged fields coupled to Dirac spinors. In Section 4 we applied the general reduction scheme to the quark sector of the Standard Model and derived quark sector Lagrangian (21). The Lagrangian is a local relativistic Lagrangian of the conventional SM quark fields interacting with the conventional SM gauge fields. It contains all the symmetries of the SM. In this section we gave the prescription for the construction of the Fock space of the theory. The main unconventional part of our theory is the construction of the Fock space. It uses an unconventional assignment of the creation and annihilation operators, as compared to their assignment in the analogous Dirac spinor theory with global $U(4)$. The assignment is forced on us by the requirement of the positivity of the Hamiltonian, the same requirement that results in the choice of the anti-commutation relations for Dirac spinors.

It is important to note that quantization causes spontaneous breakdown of $U(2,2)$, because vacuum is invariant only under $U(2) \times U(2)$ subgroup of $U(2,2)$. However, here there is no transition from false to true vacuum: all vacua are true. There is only a selection of one of inequivalent vacua by the choice of spinbeins. Since the symmetry breakdown is non-dynamical in nature, we shall refer to it as the kinematic Higgs mechanism. Since the spinbeins are physical quantities, how they have come to have particular values is an interesting question in itself. We speculate that the selection took place shortly after the Bing Bang and then was imposed on the observed part of the Universe during the inflationary phase of its expansion.

Quantization affects the fermion mass spectrum, for after quantization the matrices $\mathcal{M}_u^{AB}$, $\mathcal{M}_d^{AB}$ in (21) can no longer be absorbed in the definitions of $u$, $d$. As a result the diagonalization of the mass terms in (21) lifts $U(2,2)$ mass degeneracy and generates a non-trivial mixing matrix. For lack of space we shall give only an outline of the diagonalization procedure. Details shall be presented elsewhere.

Diagonalization of $U(2,2)$ DK action differs from its Dirac $U(4)$ analog. In the latter case one can diagonalize the mass matrix in configuration space by local $U(4)$ rotations of the fields. DK action diagonalization is local only in the phase space. The reason for the difference is that unlike the Dirac case the mass terms of the DK action contribute to kinematic part of the corresponding action. Using normalized off-shell Fourier expansion for massless spinor fields one can write the action for a single quark flavor as a bilinear form in a sixteen-dimensional Grassmann space

$$S = \int \frac{d^4k}{(2\pi)^4} C_\alpha^+ \mathcal{S}_\alpha(k,M) C_\alpha, \quad \mathcal{S}_\alpha^+(k,M) = \mathcal{S}_\alpha(k,M), \quad \mathcal{S}_2(k,M) = \mathcal{S}_1(k,M^+) \quad (28)$$

where $C_\alpha = \left(b_\alpha^A, d_\alpha^A, b_\alpha^{A+}, d_\alpha^{A+}\right)^T$, $\alpha = 1,2$. Denote $S(k,M) = \begin{pmatrix} S_{11} & S_{12} \\ S_{21} & S_{22} \end{pmatrix}$. Then action density $\mathcal{S}_1(k,M)$ for the Dirac case is

$$S_{11} = -S_{22}^* = \begin{pmatrix} gI_4 & 0 \\ 0 & gI_4 \end{pmatrix}, \quad S_{21} = \begin{pmatrix} 0 & -\mu f M^T \\ \mu f M & 0 \end{pmatrix} = S_{12}^+. \quad (29)$$

For the DK case it is

$$S_{11} = -S_{22}^* = \begin{pmatrix} gI_4 & M_c^+ \\ M_c & gI_4 \end{pmatrix}, \quad S_{21} = \begin{pmatrix} 0 & -M_d^T \\ M_d & 0 \end{pmatrix} = S_{12}^+, \quad (30)$$

$$M_d = \begin{pmatrix} \mu f \mathcal{M}_{11} & 0 \\ 0 & -\mu f^* \mathcal{M}_{22}^* \end{pmatrix}, \quad M_c = \begin{pmatrix} 0 & -\mu f^* \mathcal{M}_{12}^* \\ \mu f \mathcal{M}_{21} & 0 \end{pmatrix}, \quad M = \begin{pmatrix} \mathcal{M}_{11} & \mathcal{M}_{12} \\ \mathcal{M}_{21} & \mathcal{M}_{22} \end{pmatrix},$$

where $M = \mathcal{M}_{u,d} \in U(2,2)$ are given by (22), $\mu = m_{u,d}$, and $g = g(k^2), g^* = g$, $f = f(k^2)$ are functions of $k^2$ that depend on the spinor off-shell Fourier mode normalization. We see that the SM diagonalization

$$\begin{aligned}
\psi_L &\to V\psi_L, & \psi_R &\to U\psi_R, \\
b_2^A &\to V^{AB} b_2^B, & b_1^A &\to U^{AB} b_1^B, \\
d_1^{A+} &\to V^{AB} d_1^{B+}, & d_2^{A+} &\to U^{AB} d_2^{B+}, \\
C_1 &\to \left(Ub_1, d_1 V^+, b_1^+ U^+, V d_1^+\right)^T & C_2 &\to \left(Vb_2, d_2 U^+, b_2^+ V^+, U d_2^+\right)^T
\end{aligned} \quad (31)$$

cannot be used in the DK case. We can diagonalize either $M_d$ or $M_c$ but not both at the same time. The simplest way to determine mass spectrum is to diagonalize the DK $\mathcal{S}_\alpha(k,M)$ with a single $U(16)$ transformation $W$ and note that similar transformation applied to the Dirac case results in

$$\mathcal{S}_\alpha^D(k,M) \propto diag\left(\pm\sqrt{k^2-(m^1)^2}, \cdots, \pm\sqrt{k^2-(m^4)^2}\right). \quad (32)$$

Since the action density of a single Dirac spinor particle vanishes on-shell we can identify $m^A, A = 1,\ldots,4$ with particle masses. In the DK case one obtains

$$\mathcal{S}_\alpha^{DK}(k,M) \propto diag\left(\omega_n(k^2/\mu^2)\right), \quad n = 1,\ldots,16, \quad (33)$$

where $\omega_n(k^2/\mu^2)$ are functions that are determined by the elements of $\mathcal{M}$. Their zeroes determine the masses of the DK particle multiplet. In order to retain canonical commutation relations the DK creation and annihilation operators must be rescaled to bring $\mathcal{S}_\alpha^{DK}(k,M)$ to the same functional form as $\mathcal{S}_\alpha^D(k,M)$. Unlike in the Dirac case, the zeroes of $\omega_n(k^2/\mu^2)$ have no simple relation to the eigenvalues of $\mathcal{M}_{u,d}$, unless $\mathcal{M} \in U(2) \times U(2) \subset U(4)$. Physical states are given by zeroes of $\omega_n(k^2/\mu^2)$ at real positive $k^2$. This requirement can restrict the domain of physically acceptable parameters of $\mathcal{M}_{u,d}$. Note that our diagonalization procedure is in effect a generalized Bogoliubov transformation that is frequently used in the theory of superconductivity. We shall address this issue in more detail in a follow-up publication.

Because mass matrices (22) are different for up and down quarks, the diagonalization procedures for up and down quarks are also different. Therefore, just as in the Dirac case, diagonalization results in a non-trivial mixing matrix $V_{DK}$. Since an arbitrary $4 \times 4$ matrix $M$ has the Cartan decomposition $M = URV$ with $U, V$ unitary and $R$ diagonal positive, in the Dirac case the mixing matrix $V_D$ is given by

$$V_D = V_u V_d^+ \tag{34}$$

where $V_{u,d}$ are the $4 \times 4$ unitary matrices defined by (31) for the up and down quarks. In the DK case we can use the analogous Cartan decomposition for an arbitrary $U(2,2)$ matrix $\mathcal{M}_{u,d}$ in (23): $\mathcal{M}_{u,d} = U_{u,d} R_{u,d} V_{u,d}$, $U_{u,d}, V_{u,d} \in U(2) \times U(2)$ and

$$R_{u,d} = \begin{pmatrix} C_{u,d} & S_{u,d} \\ S_{u,d} & C_{u,d} \end{pmatrix}, \quad S_{u,d} = diag(sh\lambda_{u,d}^1, sh\lambda_{u,d}^2), \quad C_{u,d} = diag(ch\lambda_{u,d}^1, ch\lambda_{u,d}^2). \tag{35}$$

The unitary factors in the Cartan decomposition induce premixing matrix $\tilde{V}_{DK} \in U(2) \times U(2)$. Diagonalization of $R_{u,d}$ requires a Bogoliubov transformation that redefines one-particle states and induces additional mixing. If $\mathcal{M}_{u,d} \in U(2) \times U(2)$ then $V_{DK} = \tilde{V}_{DK}$ and mixing is always $2 \times 2$ block-diagonal and at tree level the first two and the second two generations do not mix. Intriguingly, the experimental data shows that the observed CKM mixing of the third generation with the first two is suppressed by roughly factor of ten [16]. The analogous PMNS mixing exhibits a different texture. However, the observed PMNS matrix can be represented as a product of two matrices, the first of which mixes only the second and the third generation in a maximal way and the second, that mixes only the first and the second generations.

It should be noted that in the SM mixing is arbitrary in the sense that it comes from the presence in the Lagrangian of all possible dimension three Yukawa coupling terms. In our case $V$ has the physical meaning of overlap mismatch of the spinbein pair $\eta, \hat{\chi}$ for the up quarks and $\eta, \hat{\theta}$ for the down quarks.

## 6. The Kinematic Higgs Mechanism and Electroweak Gauge Field Mass

Let us turn to generation of mass for gauge fields. In principle, one can use the Higgs mechanism of the Standard Model and obtain massive gauge fields this way. Then DK fermion and gauge field mass generation methods become unrelated to each other. However, DK gauge theory offers a different possibility. Instead of an additional Higgs field one can use surrogate Higgs fields obtained by combining available spinbeins. For example, consider factorizable spinbeins (20) and take $\hat{\chi} = \hat{\theta}$. Contracting $\theta$ with $\eta$, we obtain an $SU(2)$ doublet of scalars $\phi$

$$\phi_i = \mu \overline{\overline{\theta}}_i^A \eta^A = \frac{\mu}{\sqrt{2}} \lambda \varphi_i, \qquad \lambda = \sqrt{2}\, tr\mathcal{M}, \tag{36}$$

where $\mu$ is a mass parameter which is needed to make $\phi$ have dimension of mass. In the unitary gauge the vacuum state $\phi_0$ is given by

$$\phi_0^* \phi_0 = \frac{\mu^2}{2} |\lambda|^2, \tag{37}$$

where $\lambda$ is a dimensionless constant. Since in arbitrary gauge $\lambda$ can take any value, unlike $\varphi$ in (20) the doublet $\phi$ is not constrained and before gauge fixing can take any value. Using $\phi$ we can write down a gauge-invariant mass term for $SU(2)$ gauge field $W_\mu$ in the form of the Lagrangian for massless scalar doublet minimally coupled to $W_\mu$

$$\mathcal{L}_\mathcal{H} = (D^\mu \phi)^+ D_\mu \phi, \qquad D_\mu \phi = \left(\partial_\mu - ig \left(\frac{\sigma^p}{2}\right) W_\mu^p \right)\phi, \tag{38}$$

where $\phi, \lambda, \varphi$ are constrained through (36). On its own (38) describes a renormalizable interaction of massless scalar doublet $\phi$ with gauge field $W_\mu$.

In the unitary gauge for all DK fields $\Psi$, $\Phi_u, \Phi_d$ we can expand the fields around the vacuum configuration $\phi_0 = \lambda \varphi_0$, $\lambda(x) = \lambda = const$, $\varphi_{0,1} = 0$, $\varphi_{0,2} = \mu/\sqrt{2}$ to obtain the mass terms for the gauge field as the $\phi_0$-vacuum contribution to $\mathcal{L}_\mathcal{H}$

$$\Delta \mathcal{L}_\mathcal{H} = \frac{(g\lambda\mu)^2}{8} \left(\varphi_0^T \left(W_\mu^i \sigma^i\right)^+ \left(W^{k\mu} \sigma^k\right) \varphi_0 \right), \qquad \varphi_0 = \begin{pmatrix} 0 \\ 1 \end{pmatrix}, \tag{39}$$

where $\sigma^i$ are the three Pauli matrices. Lagrangian (39) is exactly the gauge field mass term that appears in the Higgs mechanism. Therefore, in the $\hat{\chi} = \hat{\theta}$ example the combination $|\lambda|\mu$ plays the same role as the vacuum expectation value $\upsilon$ of the physical Higgs field and in this example we can identify $|\lambda|\mu = \upsilon = 246\, GeV$. However, here there is no dynamical change from unstable to stable vacuum. Non-zero mass appears as a result of gauge fixing, which takes the form of a constraint on $\phi$. Hence, to describe gauge field mass generation in the DK gauge theory we shall use the term the kinematic Higgs mechanism as well. Also note that, since $\lambda$ can take any value, the mass scale $\mu$ can be considerably different from $246\, GeV$.

A more realistic is spinbein decomposition (20) with $\hat{\chi} \neq \hat{\theta}$. Now we can define two scalar doublets $\phi_u, \phi_d$

$$\phi_u = \mu \overline{\overline{\eta}}^A \chi^A = \frac{\mu}{\sqrt{2}} \lambda_u \varphi, \quad \lambda_u = \sqrt{2}\, tr\tilde{\mathcal{M}}_u,$$

$$\phi_d = \mu \overline{\overline{\theta}}^A \eta^A = \frac{\mu}{\sqrt{2}} \lambda_d \varphi, \quad \lambda_d = \sqrt{2}\, tr\mathcal{M}_d. \tag{40}$$

$\phi_u, \phi_d$ are not independent, because $\lambda_d \phi_u - \lambda_u \phi_d = 0$. In the unitary gauge

$$\phi_u^* \phi_u = \frac{\mu^2}{2} \lambda_u^2, \quad \phi_d^* \phi_d = \frac{\mu^2}{2} \lambda_d^2, \tag{41}$$

where $\lambda_u, \lambda_d$ are dimensionless constants. Now instead of (38) we can write down a gauge-invariant Lagrangian

$$\mathcal{L}_{\mathcal{H}} = \left(D^\mu \phi_u\right)^\dagger D_\mu \phi_u + \left(D^\mu \phi_d\right)^\dagger D_\mu \phi_d, \tag{42}$$

where $\phi_u, \phi_d, \lambda, \varphi$ are constrained through (41). In the unitary gauge we can choose the vacuum state $\phi_{u,1} = \phi_{d,1} = 0$, $\phi_{u,2} = \mu \lambda_u / \sqrt{2}$, $\phi_{d,2} = \mu \lambda_d / \sqrt{2}$ and obtain

$$\Delta \mathcal{L}_{\mathcal{H}} = \frac{(g\lambda_\phi \mu)^2}{8} \left(\varphi_0^T \left(W_\mu^i \sigma^i\right)^\dagger \left(W^{k\mu} \sigma^k\right) \varphi_0\right), \quad \lambda_\phi = \left(\lambda_u^2 + \lambda_d^2\right)^{\frac{1}{2}}. \tag{43}$$

We observe that when $\hat{\chi} \neq \hat{\theta}$ it is the combination $\lambda_\phi \mu$ that plays the role of Higgs' vacuum expectation value $v$. Note that, without changing the mass spectrum of the theory, we can make $\phi_u, \phi_d$ independent by using in (20) two independent spinbeins $\chi_i, \theta_i$ so that $\chi_i = \varphi_i \hat{\chi}$, $\theta_i = \tilde{\varphi}_i^* \hat{\theta}$, $\varphi_i^* \varphi_i = 1$, $\tilde{\varphi}_i^* \tilde{\varphi}_i = 1$.

The two Lagrangians (38, 42) describe only three physical degrees of freedom per surrogate Higgs doublet. For $\hat{\chi} = \hat{\theta}$ we have $\lambda = \hat{\theta}^A \eta^A = \hat{\chi}^A \eta^A$ while in the general case $\lambda_u = \hat{\chi}^A \eta^A$, $\lambda_d = \hat{\theta}^A \eta^A$. Since the spinbeins $\eta^A, \hat{\chi}^A, \hat{\theta}^A$ can always be absorbed into the definitions of the quark fields $\lambda, \lambda_u, \lambda_d$ do not represent physical degrees of freedom. Therefore, when $\hat{\chi} = \hat{\theta}$ the field $\phi$ has three physical degrees of freedom. When $\hat{\chi} \neq \hat{\theta}$ the fields $\phi_u, \phi_d$ also have three physical degrees of freedom each. However, these degrees of freedom are absorbed into the longitudinal components of the gauge field and in the end (38) or (42) do not introduce physical Higgs fields. At face value addition of either term makes the Lagrangian non-renormalizable, because in arbitrary gauge the constraints (37, 41) introduce into the combined Lagrangian an effective coupling constant that has negative mass dimension. Therefore, it should be verified whether the renormalizability of the Lagrangian is spoiled when Lagrangian terms (38) or (43) are added. In any case, in principle, such non-renormalizable terms can be considered within the framework of the effective Lagrangian approach.

## 7. Summary

We have shown that DK spinor formalism allows generation of chiral fermion mass or electroweak gauge field mass without a physical Higgs field. It naturally produces mass mixing. In addition the theory contains a discrete set of spinbein objects $\eta, \hat{\chi}, \hat{\theta}$, the symmetry properties of which might relate to the textures of CKM and PMNS mixing matrices [17].

It is interesting to compare the parameter count of the DK SM extension with that of $U(4)$ SM extension. Since the structure of the leptonic sector is not known exactly, we shall restrict ourselves to the quark sector only. The difference in the number of parameters arises from the difference in the mass terms of Dirac and DK extensions. For the Dirac case there are seventeen physical mass matrix related parameters, which are the masses of the four up quarks, four down quarks, and the nine physical parameters of the $4 \times 4$ CKM matrix. For the DK case the number of parameters is given by two mass scale parameters $m_{u,d}$ plus four non-compact parameters of $R_{u,d}$ in the Cartan decomposition (35) for up down quarks plus two physical parameters of the two premixing matrices. We obtain in total eight parameters. Therefore, the DK SM extension would have nine parameters less then its $U(4)$ extension. In fact it has less parameters then the SM.

A question arises about how the DK gauge theory fits in supersymmetry and in string theory. Supersymmetric extensions of the DK gauge theory should be straightforward. Instead of differential forms on a manifold one can use superforms on superspaces [18]. The details of such extension are unexplored. As far as string theory is concerned, a string theory with DK spinors has been known for some time [19]. Its extension to a supersymmetric form is an open question.

The DK theory postulates the existence on the fourth generation, without which it cannot be defined. The chances for discovering or ruling out the new generation at the LHC are very good, since its quark mass parameters are constrained by the precision electroweak data [9,10,20] to lie within LHC's operational range.

It also remains to be seen whether a phenomenologically acceptable DK extension of SM can be formulated that accounts for all observed experimental data. If this can be done and the fourth generation is discovered then Dirac-Kähler spinors can become an alternative to Dirac spinors for description of fermionic matter.

## Acknowledgement

I would like to thank Holger Kantz for hospitality at the Max-Planck-Institute for Physics of Complex Systems, Dresden, where a part of this research has been carried out.## References

[1] E. Kähler, Der innere Differentialkalkül, Rend. Mat. Ser., 21 (1962) 425-523.
[2] J. M. Rabin, Homology theory of lattice fermion doubling, Nucl. Phys. B201 (1982) 315-332.
[3] P. Becher and H. Joos, The Dirac-Kähler equation and fermions on the lattice, Z. Phys. C15 (1982) 343-365.
[4] T. Banks, Y. Dothan and D. Horn, Geometric fermions, Phys. Lett. B117 (1982) 413-417.
[5] I. M. Benn and R. W. Tucker, A generation model based on Kähler fermions, Phys. Lett. B117 (1982) 348-350.
[6] W. Graf, Ann. Inst. Poincare, Differential forms as spinors, Sect A29 (1978) 85-109.
[7] I. M. Benn and R. W. Tucker, Commun. Math. Phys. 89 (1983) 341-356.
[8] A. N. Jourjine, Space-time Dirac-Kähler spinors, Phys. Rev. D35 (1987) 757-758.
[9] B. Holdom *et al*, Four statements about the fourth generation, arXiv:0904.4698/v1.
[10] M. Bobrowski *et al*., How much space is left for a new family?, arXiv:0902.4883/v3.
[11] C. Quigg, Unanswered questions in electroweak theory, arXiv:hep-ph/0905.318/v1.
[12] B. Holdom, Gauged fermions from tensor fields, Nucl. Phys. B233 (1984) 413-432.
[13] D. D. Ivanenko, Yu. N. Obukhov, and S. N. Solodukhin, On antisymmetric tensor representation of the Dirac equation, Trieste Report No. IC/85/2, 1985 (unpublished).
[14] H. Grosse, P. Prešnajder, and Z. Wang, Quantum Field Theory on quantized Bergman, domain, arXiv: math-ph/1005.5723v1 and references therein.
[15] C. Itzikson and J.-B. Zuber, Quantum Field Theory, Dover, 2005.
[16] C. Amsler et al., Phys. Lett. B667 (2008) 1-1340.
[17] M. Abbas and A. Yu. Smirnov, Is the tri-bimaximal mixing accidental?, arXiv: hep-ph/1004.0099v1.
[18] T. Biswas and W. Siegel, N=2 harmonic superforms, multiplets and actions, arXiv:hep-th/0105.084v2.
[19] A. N. Jourjine, String Model with Dirac-Kähler spinors, Phys. Rev. D34 (1986) 1217-1218.
[20] G. D. Kribs, T. Plehn, M. Spannowsky and T. M. P. Tait, Phys. Rev. D76, 075016.- 12 -